  % May 16, 2002 %Oct 22,

\magnification=\magstep0
\hsize=13.5 cm               %  horizontal size of printed page
\vsize=19.0 cm               %  vertical size of printed page
\baselineskip=12 pt plus 1 pt minus 1 pt  % The line spacing
\parindent=0.5 cm  % The paragraph indent
\hoffset=1.3 cm      % The horizontal offset (may need to be changed)
\voffset=2.5 cm      % The vertical offset (may need to be changed)
\font\twelvebf=cmbx10 at 12truept % Set bold font for Title
\font\twelverm=cmr10 at 12truept % Set large font for Name
\overfullrule=0pt
\nopagenumbers    %  Actual page nos. will be inserted by the Editor
%
% The headlines
% The changes in the headlines should be made just before the Abstract
\newtoks\leftheadline \leftheadline={\hfill {\eightit Abhas Mitra}
\hfill}
\newtoks\rightheadline \rightheadline={\hfill {\eightit Nature of Compact Condensations}
 \hfill}
% Do not change the headline on the first page of paper.
\newtoks\firstheadline \firstheadline={{\eightrm Bull. Astron. Soc.
India (2002) {\eightbf 30,} 173 } \hfill}
\def\makeheadline{\vbox to 0pt{\vskip -22.5pt
\line{\vbox to 8.5 pt{}\ifnum\pageno=1\the\firstheadline\else%
\ifodd\pageno\the\rightheadline\else%
\the\leftheadline\fi\fi}\vss}\nointerlineskip}
%
% Defining 8-pt fonts for figure captions and references
\font\eightrm=cmr8  \font\eighti=cmmi8  \font\eightsy=cmsy8
\font\eightbf=cmbx8 \font\eighttt=cmtt8 \font\eightit=cmti8
\font\eightsl=cmsl8
\font\sixrm=cmr6    \font\sixi=cmmi6    \font\sixsy=cmsy6
\font\sixbf=cmbx6
%
%for switching to eight point type \eightpoint
\def\eightpoint{\def\rm{\fam0\eightrm}
\textfont0=\eightrm \scriptfont0=\sixrm \scriptscriptfont0=\fiverm
\textfont1=\eighti  \scriptfont1=\sixi  \scriptscriptfont1=\fivei
\textfont2=\eightsy \scriptfont2=\sixsy \scriptscriptfont2=\fivesy
\textfont3=\tenex   \scriptfont3=\tenex \scriptscriptfont3=\tenex
\textfont\itfam=\eightit  \def\it{\fam\itfam\eightit}%
\textfont\slfam=\eightsl  \def\sl{\fam\slfam\eightsl}%
\textfont\ttfam=\eighttt  \def\tt{\fam\ttfam\eighttt}%
\textfont\bffam=\eightbf  \scriptfont\bffam=\sixbf
\scriptscriptfont\bffam=\fivebf \def\bf{\fam\bffam\eightbf}%
\normalbaselineskip=10pt plus 0.1 pt minus 0.1 pt
\normalbaselines
\abovedisplayskip=10pt plus 2.4pt minus 7pt
\belowdisplayskip=10pt plus 2.4pt minus 7pt
\belowdisplayshortskip=5.6pt plus 2.4pt minus 3.2pt \rm}
%
% define the displayed equations to be indented 1.5 cm from left
% as required by the Bulletin. Hopefully this will work for all
% equations. With this definition using the normal $$...$$ should
% produce the equations with correct indentation, but it will be necessary
% to use \eqno to put equation numbers (though it will be possible to put
% blank eq. nos.). Further, eq. nos using \eqalignno will not work
%
\def\leftdisplay#1\eqno#2$${\line{\indent\indent\indent%
$\displaystyle{#1}$\hfil #2}$$}
\everydisplay{\leftdisplay}
%
% Some useful definitions
% less than or order of \la
\def\frac#1#2{{#1\over#2}}

% greater than or order of \ga

%
%
%to generate boldface characters
\def\pmb#1{\setbox0=\hbox{$#1$}\kern-0.015em\copy0\kern-\wd0%
\kern0.03em\copy0\kern-\wd0\kern-0.015em\raise0.03em\box0}
%
%Beginning of Document%
\pageno=1
\vglue 60 pt  %Leave some space on page 1 before the title
% The title
%

\leftline{\twelvebf On the nature of the compact condensations at the centre of galaxies}

% if more than one line is required for the title, then use next two lines ...
%
\smallskip
% end of title
\vskip 46 pt  % Space between title and author(s) name(s).

\leftline{\twelverm Abhas Mitra} % Name of Authors
\vskip 4 pt
\leftline{\eightit Theoretical Physics Division, BARC, Mumbai- 400085
India (amitra@apsara.barc.ernet.in).}

%\leftline{\eightit  to reduce the number of lines}
%
% If authors are from different institutes, repeat the above lines
% for each institution. For authors from same institution write the
% names in one line.
%
%\vskip 0.5 cm
%\leftline{\twelverm V. R. Co-author1 and V. R. Co-author2}
%\vskip 4 pt
%\leftline{\eightit Name and Address of the institution}
\vskip 20 pt % leave some space between author(s) names(s) and abstract
%
%
% The leftheadline should include the Authors' name, for two authors use
% \&  (e.g. I. M. Author \& I. M. Co-author) for three or more authors
% use et al.,
\leftheadline={\hfill {\eightit A. Mitra} \hfill}
% Use a short running title as the rightheadline
\rightheadline={\hfill {\eightit  Nature of Compact Condensations
}  \hfill}

% Abstract begins
%
{\parindent=0cm\leftskip=1.5 cm

{\bf Abstract}
There are many observational evidences for the existence of massive
compact condensations in the range $10^6 -10^{10} M_\odot$ at the core of
various galaxies and in particular in the core of High Energy Gamma Ray
emitting galaxies.  At present such condensations are commonly interpreted
as Black Holes (BHs). However, we point out that while such Black Hole
Candidates (BHCs) must be similar to BHs in many respects they, actually,
can not be BHs because existence of Black Holes would violate the basic
tenet of the General Theory of Relativity (GTR) that the {\it worldline of a
material particle must be timelike at any regular region of spacetime}. On
the other hand general relativistic collapse of massive bodies should lead
to Eternally Collapsing Configurations (ECOs). While ECOs may practically
be as compact as corresponding BHs, they will have a physical surface.
Also while BHs do not have any intrinsic magnetic field ECOs may have
strong intrinsic magnetic field.  We point out that despite many claims
actually there is no real evidence for the ``Event Horizon'' (EH) of supposed
BHs and on the other hand, there are tentative evidence for the existence
of strong magnetic field in several BHCs (or ECOs).  The presence of such
intrinsic magnetic field may render the task of explaining high energy
radiation phenomenon in many Active Galactic Nuclei easier.
%\footnote{Invited talk given in the GAME 2001 Conference, Mt. Abu, India, 2001}
\noindent
\smallskip
\vskip 0.5 cm  %  Space between Abstract and Key words
{\it Key words:} Compact Objects, No Black Holes

}                                 %  End of abstract
% Beginning of document
%
%
% Beginning of a section heading
%
% for the first section leave 20 pt space, for subsequent sections just
% leave bigskip (i.e. 12 pt)
\vskip 20 pt
\centerline{{\bf 1. Introduction}}
\bigskip
This conference GAME 2001 is aimed at understanding the High Energy
Astrophysics associated with gamma ray sources and for the specific case
of point sources it is important to know the nature of the central engine
accelerating the particles responsible for gamma production. One of the
rules of this GAME so far has been to call all cold compact objects having
mass higher than 3-4 $M_\odot$  as ``Black Holes'' (BHs). But we would
appeal to change this rule. While we make this appeal, we must say that, in
the existing paradigm,
 there are, apparently, many good reasons to believe that all such massive
compact condensations are BHs or singularities. We know that all stars do
exhaust their nuclear fuel at a certain finite time and therefore must
start collapsing due to self-gravity. Low mass stars, following
gravitational collapse, end up as a class of compact objects called White Dwarfs
(WDs) where the gravitational pull is counteracted by degenerate electron
pressure of the stellar material. But Chandrasekhar taught us that there
is an upper limit
on the mass of WDs (which, at the time of its proposition, sounded as incredible
as the present appeal for a new ``rule''). This implied that more massive
stars, at the exhaustion of their fuel, must collapse to a stage beyond
the WD one. Now we know that more massive stars may collapse to deeper
potential wells supported by the degeneracy pressure of their nucleons.
Such configurations are broadly referred to as Neutron Stars (NSs)
although such configurations could be Strange Stars or other cold baryonic condensations.
We also know that NSs do have an upper limit on their masses and in the
context of {\it standard} Quantum Chromodynamics, an absolute upper limit could be
$M_{ov} \sim 4-5 M_\odot$. This again means that very massive stars must
collapse to a stage beyond the NSs stage which is generally called as BHs.
 Atleast for the spherical case, this seems to be obvious
because of the following (incorrect) thinking: Let a given star has initial {\it radius
parameter} $R_i$. Then suppose the star collapses with a mean local speed
$V_m$. Then, as a matter of geometry, it might appear that, the star would
collapse to a geometrical point, a singularity, in a comoving proper time
$R_i/V_m$. While this is true in Newtonian gravitation, this picture is
actually not strictly correct in GTR because {\it $R_i$ is not the
physical radial depth of the star} eventhough $2\pi R_i$ is its circumference! If $r$ is the comoving radial
coordinate, and $g_{rr}$ is the corresponding metric coefficient (see
below), the physical or proper or locally measured radial depth of the
star is somewhat like
$l \sim \int_0^{R} \sqrt{-g_{rr}} dr$.
In a dynamical case $l$ is not properly defined because $g_{rr}$ is ever changing,
yet one can obtain a definite value for $l$ in
the limit $R \rightarrow 0$. Since $- g_{rr} \rightarrow \infty$ in the same
limit, it is probable that $l$ might blow up. And we have
found that this is indeed the case (Mitra 1998, 2000). As the star tries to collapse to
deeper and deeper potential well, the grip of gravity stretches the
physical space more and more. And as, eventually, the space stretching
tidal gravity (components of the Ricci Tensor) tends to become infinite,
{\it the inner or physical radial space becomes infinite too}!.  Since by
principles of relativity $V_m$ is always finite the collapse never
terminates (in a finite proper time)! As an
oberver sitting on the surface of the star tries to chase its centre, the
chase becomes longer and longer like a Xeno's paradox.
 The observer never reaches that ``geometrical point''
and simultaneously he too as a participant in this spacetime game becomes
of infinite proper length. Now compare this with the popular folk lore of
an observer/``astronaut'' falling into a supposed BH or consider the
situation of a collapse when the star would indeed become a (finite mass)
BH. Here the observer is supposed to reach and get crushed in a geometrical
point in a finite proper time. But here too, the proper length of the
observer becomes infinite. But how does the {\it observer of infinite
proper length stay put in a point}? There is no answer to such ludicrous
incongruities in the BH paradigm and a self-consistent answer can be found
only in terms an ECO. Technically, this implies that, {\it
for isolated bodies} GTR is singularity free! Since our discussion
specifically uses isolated bodies with definite boundary condition, this
result cannot be extended to the Universe or in other words, our work does
not rule out the ``Big Bang'' singularity.

 In order to appreciate our result, at the very outset, it is necessary
to distinguish between the concepts of {\it gravitational mass} and
baryonic mass. The mass-energy of a nucleus is less than the sum of the
masses of the individual nucleons because of the (negative) attractive
nucler Binding Energy (or Mass Defect). Similarly, in the simplest case, the net mass energy or
the gravitational mass of a star $M$ (as perceived by a distant observer) is
less than the sum of rest masses of its individual baryons ($M_0$): $M= M_0
- B.E.$. Long back, Zeldovich and Novikov (1971) conceived of a tightly packed
self-gravitating configurations of baryons where the $B.E. = M_0$.
Obviously, in such a case, the gravitational mass of the system would be zero
$M=0$ ! Even much before this Harrison et al. (1965) postulated that for a system
of self-gravitating baryons, {\it there exists a final state where} $M=0$!
To appreciate such ideas first one has to understand that as a
self-gravitating system undergoes gravitational compression, it emits
radiation and therefore the value of $M$ keeps on decreasing. Now during
collapse, the value of $R$ of course decreases and $R\rightarrow 0$. But
how does the ratio, $\alpha= 2GM/R c^2$ would change during this process. In
Newtonian gravity, $M=M_i$ is constant and $\alpha$ increases relentlessly;
for $M= 1 M_\odot$, $\alpha$ becomes unity at $R= 3$ km, and for any value of
$M$ eventually $\alpha = \infty$ at $R=0$. Since our common sense is
governed by Newtonian physics, in the context of GTR too, we take it for
granted that $\alpha$ would behave in more or less the same (Newtonian)
 fashion even though $M$ may be changing. It is this Newtonian common sense
by which the assumption of formation of a ``trapped surface'' during a GTR
collapse is taken for granted. Nevertheless, we have found that {\it this
common sense} is incorrect and GR does not allow existence of trapped
surfaces and BHs (Mitra 2000). While we
say this, it may seem that the present speaker is an absolute loner in this respect. But it
is not so:

First it is well known that initially most of the founder fathers of GR
considered the idea of a BH to be unphysical, some of the prominent names here are
Schwarzschild, Weyl, Eddington, Rosen and most importantly Einstein (1939)
himself.
Interstingly and surprisingly, the conventional GR solution for the
spacetime around a point mass, which is supposed to consolidate the
concept of a BH, is not due to Schwarzschild! On the other hand this
conventional solution is due to Hilbert (point mass sitting at $R=0$)! In
the original Schwarzshild solution (whose English translation has recently
been made by Antoci and Loinger (1999)), in a trivial manner, there is no EH
and no BH, because here the ``point mass'' is sitting {\it not} at $R=0$, but at
$R_* =0$ where
$$R = [R_*^3 + ( 2GM/c^2)^3]^{1/3} \eqno(1)$$
The EH $R= R_g= 2GM/c^2$ corresponds, in original Sch. picture, to $R_*=0$, the
origin of the coordinate system. Thus there is no finite EH and no
spacetime beneath $R=2GM/c^2$. Following this cue, Loinger (Univ. of
Milan) (1999, 2000), Antoci and Leibscher (Univ. of Padavo, Italy) (2001) ,
 Zakir (Tashkent) (1999), Abrams (Canada) (1989)
have exterted that there are no BHs in GR. On the other hand,  Leiter and
Robertson (Univ. of South Okalohoma) (2001), on the basis of a modified form of
GR, have shown that there may not be BHs. But my approach to the problem
has been quite different. In the following, let me sketch
the reasons why there can not be any finite mass BHs (henceforth $G=1$).

\bigskip

\centerline\bf{2. Test Particle Radially Falling on a BH}
\bigskip
\noindent We know from Sp. Relativity, that for any event, the associated spatial
coordinate ($\vec {R}$)
and time (t) can be amalgamated as a spacetime interval in a fictitious
4-D spacetime ($s^2 = c^2 t^2 - {\vec{R}}^2$). An infinetisimal interval ``metric''
is $ds^2 = c^2 dt^2 - d{\vec {R}}^2$. In a spherically symmetric case, for
a particle on a radial motion, ${\vec R} = R$ and $ds^2 = c^2 dt^2 - dR^2$.
The foundation of Sp. relativity lies on the tenet that for a material particle
$ds^2 > 0$ (Worldline is Timelike) and for massless particles like photons
$ds^2 =0$ (Worldline is Null). The above tenets are equivalent to the
more popular tenet that ``the speed of a photon is always fixed $V=c$
while the former condition means that ``the speed of a material particle
is always less than $c$''. This is so because, it can be seen that
$$ds^2 = c^2 dt^2 [ 1 - V^2/c^2] \eqno(2)$$
where the speed of the particle as measured by the given observer is $V= dR/dt$
(for the radial case). Thus $ds^2 >0 $ implies $ V <c$ and $ds^2 =0$
implies $V=c$.
 Since, by Principle of Equivalence, locally, GR reduces to Sp. Relativity
 even within a supposed BH or EH, in GR too, for a material
 particle,
 always, $ds^2 >0$ and locally measured
speed of a material particle, as measured by any observer, must be less
than $c$ (at a singularity, it is possible that $ds^2=0$ and $V=c$).
\bigskip
\centerline{\bf {3. Spherically Symmetric Gravitational Field}}
\bigskip
Any spherical gravitational field may be expressed, in terms of general
time coordinate $x_0$ and radial coordinate $x_1$ as
$$ds^2 =   g_{00} dx_0^2 +  g_{11} d x_1^2 -  R^2 (d\theta^2 +\sin^2
\theta d\phi^2) \eqno(3)$$
Here $R$ is the {\it Invariant} Circumference Coordinate (a scalar).
Since, we shall deal with only radial worldlines with $d\theta = d\phi=0$,
our effective metric will be
$$ds^2 =   g_{00} d{x^0}^2 +  g_{11} d {x^1}^2 \eqno(4)$$
Following Landau \& Lifshitz (1985), any general metric involving $x^0$
(time coordinate) and $x^1$ (radial coordinate)  can be rewritten  as
$$ds^2 =   g_{00} d{x^0}^2 [1- V^2] \eqno(5)$$
where
$$V\equiv {\sqrt{-g_{11}} d x^1\over \sqrt{g_{00}} dx^0}= dl/d \tau \eqno(6)$$.
Here we take speed of light $c=1$ (also $G=1$) and $ d\tau$ is element of proper time.
It is easy to see that Eq.(5) is the GTR generalization of Eq.(2). If
there are spacetime cross terms (i.e, rotation) in Eq.(3), Eqs.(5) and (6)
needs to
be modified, but here we are not interested in such a case.

For the spacetime around a  Sch. BH (SBH) (actually Hilbert BH),
the radial variable happens to be same as $R$, i.e, $x^1 \equiv R$. It is
a known result that in this coordinate, the speed of a test particle on
the EH becomes exactly equal to the speed of light, $V=c=1$. Had Eq.(5)
been logically
pursued in the past, it would have been taken as a pointer for the non
existence of BHs because as my previous speaker correctly exerted
(Chakrabarti 2001), the speed of a test particle on the EH must be $V=c=1$.
 One natural question here could have been, ``if the
speed becomes already $c$ at the EH  will the speed not exceed $c$
once the particle crosses it?'' But  such physical questions are a taboo as
far as modern research in GTR is concerned and  such a
questioner would be seen as
an odd man out, a person cutoff from the profound and modern BH research
(though by principle of equivalence GR always reduces to Sp. Rel. locally, in a
free falling frame, and one can should always be able to deal with
physical quantities like speed, acceleration and differential acceleration
or components of Ricci Tensor). There are several apparent reasons
and excuses to not to even pose this question. Some of scientific excuses
amongst them  could be, (i) The
external Sch. coordinate system breaks down inside the EH and (ii) the
geodesic, nevertheless remains timelike even when the particle is on the EH. As to the second assertion,
recall that, for the external Sch. metric, $g_{00}= 1- 2M/R= 0$ on the EH,
further $d x_0$ is an infinetisimal (not $\infty$) and therefore, one can
see from Eq. (5) that as $V=1$ $ds^2 =0$. In other words the second
assertion is incorrect on the EH.  We have obtained this conclusion even
without introducing any $V$ at all in the problem (Mitra 2000, 2001).

 To tackle the former argument, one can move on to the Kruskal coordinates
  which is believed to correctly describe the entire spacetime associated
with a SBH. Here again, it has been found that, the speed of a test particle
as measured by Kruskal coordinates becomes equal to the speed of light.
Further, using the form of the metric in terms of Kruskal coordinate, it
has been shown
 that the geodesic associated with its motion becomes null at $R=2M$
(Mitra 2000, 2001).
It must be so because value of $ds^2$ is independent of the coordinates used.
Although, now, the {\it belivers} in BH hypothesis really have no argument left to
justify their faith, most of them would just tend to ignore and forget
the whole matter quietly!
There is an interesting physical reason why the validity of the result
$V=c$ (in Sch. coordinates) cannot be shrugged off with the pretext of a
``coordinate singularity'' on the EH: Recall the Sp. Rel. velocity addition law
for two speeds: $V = (V_1 + V_2)/(1+ V_1 V_2)$. Once either or both of
$V_1$ and $V_2 \rightarrow 1$, it follows that the resultant velocity $V
\rightarrow 1$. Thus once a given observer perceives a speed to be equal
to $c$ all other observers too perceive it to be $c$! Although, in GTR,
velocity addition law is different,
this basic result remains unchanged there and this is the reason that  in
GTR too, {\it all observers measure the same speed for light} and nothing
can exceed it.
The physical implication of the fact that $ds^2 =0$ (rather than $>0$) at
$R=2M$ is that the EH is the true singularity and not a mere ``coordinate singularity''.
The only true singularity in the problem is, however, the central singularity.
These two singularites merge when the radius of the EH is zero, i.e,
$2M=0$. Thus if one insists for a BH, mathematically its mass must be
$M=0$, a possibility  considered previously by many authors.

 There is yet another direct proof that the EH is the true central singularity.
Like 4-velocity $u^i = dx^i/ds$, one can also define 4-acceleration $a^i$.
The norm of any 4-vector is a scalar and must be non-singular at a mere
coordinate singularity (the norm of $u^i$ is $u= \sqrt {u^i u_i} = c$. One
finds that the norm of 4-accel. is (Abrama 1989, Antoci and Leibscher
2001, Mitra 2001)
$$a = \sqrt {a^i a_i} = {M \over R^2 \sqrt{1 - 2M/R}}\eqno (7)$$
Note that not only does $a$ blow up at the EH $R= 2M$, had there been a
spacetime beneath the EH, $a$ would have become imaginary. $a$ being a
scalar is coordinate independent and in this case measurable. This doubly
confirms our result that there can be no EH and no BH unless its mass is
trivially zero. And when we realize that in GR, for {\it isolated
singularities} or ``point masses'', one must have $M=0$, we see that the
original Sch. solution becomes synonymous with the conventional Sch.
(Hilbert) solution (Mitra 2001).
\bigskip
\centerline{\bf{4. Spherical Gravitational Collapse}}
\bigskip
\noindent It follows from the most general formalism of spherical
gravitational collapse (Mitra 2000 and ref. therein) that the
integration of the $0,0$ component of the Einstein equation leads to a constraint
$$\Gamma^2 = 1 + U^2 - 2 G M(r)/R \eqno(8)$$
where  $M(r)$ is the gravitational
mass enclosed by a shell with $r=r$. Here the parameters
$\Gamma= {dR\over d l};~~
U = {dR\over d \tau}  $,
so that, by using Eq.(6), we have
$U=\Gamma V $
By inserting this relation in Eq.(8), and by transposing, we find,
$$\Gamma^2 (1- V^2) = 1 - 2 G M(r)/R \eqno(9)$$
Now it follows from Eq.(5) that if $1-V^2$ is considered positive so is
$\Gamma^2$, but if it is assumed to be negative again so will be
$\Gamma^2$ (Mitra 2001). This means that whether we interpret $V$ as the local speed or
not, the LHS of the above Eq. is always positive, so that
$${2G M(r)\over R} \le 1 ; \qquad {R_{g} \over R} \le 1 \eqno(10)$$
On the other hand, the condition for formation of a ``trapped surface'' is
that $2G M/R >1$. Thus we find that, in spherical gravitational collapse
 {\it trapped surfaces do not form}.
If the collapse process  indeed continues upto $R=0$, in order that the
foregoing constraint is satisfied, we must have
$M(r) \rightarrow 0$ as $ R\rightarrow 0$, a conclusion we have already
obtained from several considerations (Mitra 2000).
It is widely believed that by studying the problem of the collapse of the
most idealized fluid, i.e, a ``dust'' with pressure $p \equiv 0$ and no
density gradient, Oppenheimer and Snyder (OS) (1939) explicitly showed that
finite mass BHs can be generated. We have discussed in detail (Mitra 2000) that
this perception is completely incorrect. For the sake of brevity, we would
like to mention here about the Eq.(36) of  OS paper which connects the
proper time $T$ of a distant observer with a parameter
$y= {R\over 2 G M}$ (at the boundary $r=r_b$) through the Eq.
$$T \sim \ln {y^{1/2} +1\over y^{1/2} -1} + ~~ other ~ terms.\eqno(11)$$
In order that $T$ is definable,  the argument of this logarithmic term must
be non-negative, i.e,
$y= {R\over 2 G M} \ge 1$, or,
${2GM\over R} \le 1$,
which is nothing but our Eq.(9). Thus even for the most idealized cases,
trapped surfaces are not formed. Hence there are no BHs (of finite mass).
 However, one can still legimately wonder, if one starts with a dust of finite
 mass $M$, {\it and if the dust does not radiate}, why the condition
$2GM/R >1$ would not be satisfied at appropriate time? The point is that
since for a dust $p=0$, dust is really not a fluid, a (spherical) dust is just a collection of
incoherent finite number of $N$ particles distributed symmetrically. If
so, there are free spaces in between the dust particles and which is not the
case for a ``continuous'' fluid. Therefore although
the dust particles are symmetrically distributed around the centre of
symmetry, in a strict sense, the distribution is not really isotropic. Then the assembly of
{\it incoherent} dust particles may be considered as a collection of $N/2$
symmetric pair of particles. In GTR, a pair of particles accelerate each
other and generate gravitational radiation unmindful of the presence of
other {\it incoherent} pairs. Therefore, the gravitational mass of an
accelerating dust is really not constant! In contrast a physical spherical fluid
will behave like a coherent single body with zero quadrupole moment and
will not emit any gravitational radiation.

\bigskip
\centerline{\bf{5. Summary and Discussions}}
\bigskip
From several independet approaches and from most basic premises, our work has
shown that GTR does not allow existence of BHs. The immediate question
would be then what is the true nature of the observationally discovered BH
candidates (BHC). Our work suggests that continued collapse of very massive
objects continues indefinitely because
{\it the ever incresing curvature of spacetime (Ricci Tensor) tends to stretch
the physical spacetime to infinite extent}. Hence such objects have been called
Eternally Collapsing Objects (ECO).
 If in a given epoch the gravitational
mass of a BHC/ECO is $M$, its circumfernce radius $R$ can be arbitrary close
to its Sch. radius $R_g= 2GM/c^2$ without ever becoming less than $R_g$,
i.e, $R \ge R_g$. For practical purposes such objects are as compact as a
supposed BH and they would satisfy many of the ``operational definitions''
of BHs. For instance, my previous speaker (Chakrabarti 2001) asserted that Galactic BHCs show
a hard power law X-ray component going well beyond 100 keV. He explained
that such a hard X-ray tail may be understood as direct Compton
upscattering of seed phtons by the electrons of the relativistically
moving plasma accreting on a BHC. He pointed out that for accretion onto a NS,
the speed of the plasma is smaller, the electrons are less energetic, and
such a hard power law tail cannot be expected there. For a NS accretion,
the maximum speed of the plasma is $V \approx 0.5 c$ and the corresponding
Lorentz factor is paltry $\sim 1.1$. So the plasma is hardly relativistic
and indeed such a hard power law may not be generated there (by the
assumed process). On the other hand, Chakrabari requires a value of
$\gamma \sim 1.5 -1.6$ for generating the hard X-rays. Recall here that
the EH is characterized by a gravitational redshift $z=\infty$ and matter
falling on it acquires
a value of $\gamma = \infty$. Also recall here
that if the gravitational surface redshift of the compact object is $z$,
then for free fall of plasma, one has $\gamma = 1+z$.
So assuming that
Chakrarbati's model is indeed correct, the observation of a hard X-ray tail does not
necessarily prove the existence of BHs (i.e, $z=\infty$),
 but, on the other hand, it simply indicates the existence of objects
whose value of $z$ is reasonably above $0.6$. So an ECO with $z <2$ should
be sufficient to explain this observation.
Such BHCs or ECOs with finite $z$  need not always be
static and cold, they need not represent stable solutions of equations for
hydrostatic balance.  They may be
collapsing  with substantial local speed $V$, (which this work cannot predict)
 but the speed of collapse
perceived by a distant observer ($V_\infty$) would approximately be lower by a factor
of $(1+z)^2$.  Since $V$ is finite ($<c$) and eventually $z \rightarrow
\infty$, the ultimate value of $V_\infty \rightarrow 0$. So it is likely that
even for accurate measurements (which might be possible in remote future)
spanning few years, an {\it isolated} ECO may appear as a ``static'' object.
The value of $R$ for an accreting ECO would decrease even more slowly.
 Thus gravitational collapse of sufficiently massive bodies should indeed
result in objects which could be more compact than typical NSs ($z > \sim 0.1)$.
It is found that, if there are anisotropies, in principle there could be
static objects with arbitrary high (but finite) $z$ (Dev and Gleiser 2000). Even within the
assumption of spherical symmetry, non-standard QCD may allow existence of
cold compact objects with masses as large as $10 M_\odot$ or higher
(Miller et al.  1998).
Such stars are
called Q-stars (not the usual quark stars), and they could be much more compact than
a canonical NS; for instance,
a stable  non rotating Q-star of mass 12$ M_\odot$ might have a radius of
$\sim 52$ Km. This may be compared with the value of $R_{gb} \approx
36$ Km of a supposed BH of same mass.
 And, in any
case, when we do away with the assumption of ``cold'' objects and more importantly,
staticity condition there could be objects with arbitrary high $z$.

   There is another line of argument for the having found the existence of
EHs in some BHCs (Garcia et al. 2001). At very low accretion rates, the coupling between
electrons and ions could be very weak. In such a case most of the energy
of the flow lies with the ions, but since radiative efficiency of ions is
very poor, a spherical flow radiates insignificant fraction of accretion energy
and carries most of the energy towards the central compact object. Such a
flow is called Advection Dominated Flow (ADAF). If the central object has
a ``hard surface'', the inflow energy is eventually radiated from the hard
surface. On the other hand, if the central object is a BH, the flow energy
simply disappears inside the EH. For several supermassive BHCs and stellar
mass BHCs, this is claimed to be the case.
But   there could be several caveats in this interpretation:

 (i)  The observed
X-ray luminosities for such cases are usually insignificant compared to
the corresponding Eddington values (by a factor $10^{-5}$ to $10^{-7}$).
Such low luminosities may not be due to accretion at all. Atleast in some
cases, they may be due to Synchrotron emission.
Recently, Robertson (1999) and Robertson and Leiter (2001) have attempted to explain the X-ray
emission from several BHCs having even much higher luminosities as
Synchrotron origin. Vadawale, Rao and Chakrabarti (2001) have explained one
additional component of hard X-rays from the micro-quasar GRS1915+105 as
Synchrotron radiation. The centre of our galaxy is harbours a BHC, Sgr A$^*$, of mass
$2.6 \times 10^6 M_\odot$. The recent observation of $\sim 10-20\%$ linear polarization
from this source has strongly suggested against ADAF model (Agol 2000). On the other
hand, the observed radiation is much more likely due to Synchrotron
process (Agol 2000).
In fact, even more recently, Donato, Ghisellini \& Tagliaferri (2001) have shown that
the low power X-ray emission from the AGNs are due to Synchrotron process rather
by accretion process.

(ii) The x-rays if assumed to be of accretion origin, could be coming from
an accretion disk and not from a spherical flow.

(iii) Even if the X-rays are due to a spherical accretion flow, not in a
single case, we have robust independent estimate of the precise accretion rate.

(iv) Munyaneza \& Viollier (2001) have claimed that the accurate studies of the
motion of stars near Sgr A$^*$ are more amenable to a scenario where it is
not a BH but a self-gravitating ball of Weakly Interacting Fermions of
mass $m_f >\sim 15.9$ keV. Recall here that the Oppenheimer - Volkoff mass
limit may be expressed as
$$M_{OV} = 0.54195 M_{pl}^3 m_f^{-2} g_f^{-1/2} = 2.7821 \times 10^9 M_\odot
(15 keV/ m_f)^2 (2/g_f)^2 \eqno(12)$$
where $M_{Pl}= (\hbar c/G)^{1/2} $ is the Planck mass and $g_f$ is the
degeneracy factor. With a range of  $ 13<m_f <17 $ keV, these authors
point out that the entire range of supermassive BHCs can be understood.
Bilic (2001) has also suggested that the BHCs at the centre of galaxies
could be heavy neutrino stars. Svidzinsky (2001) has suggested that atleast some
of the BHCs in the blazars could be heavy bosonic stars.
Note that the progenitors of the ECOs or BHCs must be much more massive (and larger in size)
than those of
the NSs.
 Then it
 follows from the magnetic flux conservation law that BHCs (at the galactic level)
 should have
magnetic fields considerably higher than NSs. It is also probable even
when they are old, their diminished magnetic fields are considerably
higher than $10^{10}$G. In such cases, BHCs will not exhibit Type I X-ray
burst activity. There may indeed be evidence for
 intrinsic (high) magnetic fields for the BHCs (Roberson 1999, Robertson
and Leiter 2001). However, in
some cases, they may well have sufficiently low magnetic field and show
Type I bursts. It is now known that Cir X-1 which was considered a BHC,
did show Type I burst, a signature of ``hard surface''. Irrespective of
interprtation of presently available observations, our work has shown that the
BHCs can not be, in a strict sense, (finite mass) BHs because then {\it timelike
geodesics would become null} on their EHs.

 Starting from GRB 971214 there are several powerful Gamma Ray Bursts
which show no evidence for beamed emission. For GRB 971214, the total
power radiated only in soft gamma rays is $Q\sim 3 \times 10^{53}$ ergs.
Since the efficiency for gamma production could be considerably below
100$\%$ and since there could be substantial associated neutrino emission,
the actual energy liberated in such bursts could be well above $10^{54}$
erg. If gravitational collapse of massive stars would have resulted in
prompt formation of trapped surfaces, such huge energy emission
 would have been nearly impossible.

 Explanation of many high energy phenomenon reqires postulation of a
quasi-spherical {\it
standing shock} around the BHCs. When the BHCs have a physical surface and
an intrinsic magnetic field it is easy to understand the formation of such
{\it standing} shocks. On the other hand, it is extremely difficult to
conceive of a {\it standing} shock supported by the EH, i.e, by {\it no
physical surface} just like it is not possible to have a stable floor for a
building which has no foundation at all.
 We know for certain that the astrophysical systems like NSs and young
protostars possessing physical surface, intrinsic magnetic and rotation do
allow formation of ``jets''. However an impression often is created in the
literature that in order to explain jets from gamma sources and AGNs, it is
necessary to assume the existence of BHs.  The fact is that while
accretion disks (physical surfaces with magnetic field)
around any object might be site for a jet formation, it is
not understood how jets can emanate from a BH which gulps up everything
 or how exactly a BH helps
formation of jets. For a NS with weak magnetic field, the inner disk of
the accretion disk extends upto the stellar surface in order to find a
``support'' from a physical surface. Similarly, for a BH, the accretion disk
should tend to extend all the way upto the central singularity, if it were
possible, in the absence of any physical surface even though stable
keplarian circular orbit is possible only upto $R= 3 R_g$. Then, can there
really be reasonably stable accretion disks around BHCs if they were BHs? For BH
accretion, in the absence of a physical surface it is extremely difficult
to see how the accretion flow can get ``rebounded'' unless the jet is
launched far away from the disk. All such conceptual problems are absent
when we realize that BHCs are actually ECOs which in a broad sense may
behave something like a magnetized NS. Then it becomes much easier to
understand acceleration of charged particles by the BHCs in the AGNs.
It is one of the tasks for
Gamma Ray Astronomy to eventually confront such questions and unravel the
scientific truth.

% Sample references
\bigskip
\centerline{\bf References}
\bigskip
{\eightpoint\parindent=0pt\everypar={\hangindent=0.5 cm}
% References in the format of the Bulletin of the Astronomical Society of India
% using 8 pt fonts
% leave one line blank between two references to force a paragraph break
%scussions.
  Abrams, L.S., 1989,  Can. J. Phys.,  67, 919.

 Agol, E, 2001,  Astrophys. J. Lett., (in press), astro-ph/0005051.

  Antoci, S. and  Liebscher, D.E., 2001, gr-qc/0102084.

  Antoci, S. and  Loinger, A., 1999, physics/9905030.

Bilic, N., 2001, (astro-ph/0106209).

 Chakrabarti, S.K., 2001, This volume.

% Chou, W. and  Tajima, T., 1999, Astrophys. J., 513, 401.

  Dev, K. and  Gleiser, M.  2000, astro-ph/0012265.

   Dotani, D.  Ghisellini, G,  Tagliaferri, G, 2001,  Astron.
Astrophys. (in press), astro-ph/0105203.

  Einstein, A., 1939,  Ann. Math., 40, 922.

 Garcia, M.R.,  McClintock, J.E.,  Narayanan, R. and Callanar, P., 2001,
 Astrophys. J., 553, L47, astro-ph/0012452.

Harrison, B.K., Thorne, K.S., Wakano, M. and Wheeler, J.A., 1965, in
Gravitation Theory and Gravitational Collapse (University Chiago,
Chicago), p. 75

 Landau, L.D. and  Lifshitz, E.M., 1985,  The Classical Theory of
Fields, 4th ed., (Pergamon Press, Oxford).

  Leiter, D.J. and  Robertson, S.L., 2001, General Relativity and
Gravitation (submitted), gr-qc/0101025.

  Loinger, A., 2000, astro-ph/0001453, also, 1999, gr-qc/9908009.

  Miller, J.C.,  Shahbaz, T. and  Nolan, L.A., 1998,
 MNRAS, 294, L25.

  Mitra, A., 1998 (astro-ph/9803014).

  Mitra, A., 2000,  Found. Phys. Lett.,  13, No. 2, 543.

Mitra, A., 2001, Indian J. Phys. (in press),
Invited talk given in Black Hole Workshop, Calcutta,
2001 (astro-ph/0105532).

Munyaneza, F. and Viollier, R.D. 2001,  Astrophys. J. (submitted),
 astro-ph/0103466.

  Oppenheimer, J.R. and  Snyder, H., 1939, Phys. Rev., 56, 455.

 Robertson, S.L. and Leiter, D., 2001, Astrophys. J. 565, 447,
  astro-ph/0102381.

 Robertson, S.L., 1999, Astrophys. J., 515, 365.

Svidzinsky, A., 2001 (astro-ph/0105544).

Vadawale, S.V.,  Rao, A.R. and  Chakrabarti, S.K., 2001,  Astron.
Astrophys,  (submitted), astro-ph/0104378.

  Zakir, Z., 1999, gr-qc/9905068.

Zeldovich, Ya. B. and Novikov, I.D., 1971, in Relativistic Astrophysis,
Vol. I, p. 297 (Univ. Chicago, Chicago).
}
\bigskip
Note added after publication: Having sent this manuscript to press, we became aware of a preprint
(gr-qc/0109035, Phy. Rev. Lett., submitted) entitled ``Gravitational Condensate Stars: An Alternative
to Black Holes'' by P.O. Mazur and E. Mottola. These authors have suggested that immediately before formation
of Event Horizon, quantum effects arising due to extremely strong gravity may cause a phase transition
 of the collapsing matter form $p =- \rho$, the collapse may be halted
and there could be static Ultra Compact Objects of arbitrarily high mass.  By using the theory
of General Relativistic Polytropes, the present author has also found that even if there is a much
more modest (causality obeying) phase transition of the form $p \rightarrow \rho$, there could be
static UCOs of arbitrary high mass, and, in particular, if such a phase transition would occur
at $\rho =\rho_{nclear} \sim 2. 10^{14}$ g/cm$^3$, the maximum mass of such a configuration would
be $\sim 11 M_\odot$ (A. Mitra, in preparation).

\bigskip
For previous instance of use of the concept of local 3-speed even when one is
working with comoving coordinates see

(i) Eqs. 2.79-80 in ``An introduction to mathematical cosmology''

by  J.N. Islam, Cambridge Univ. Press (1992).

and

(ii) Eqs. 6.42-45 in ``Cosmolgy and Astrophysics through problems''

by T. Padmanabhan, Cambridge Univ. Press (1996).

% End of section heading
%\endref
                                         % End of references
% leave one line blank before the closing braces.
\end